\RequirePackage{lineno}
\documentclass[preprint,showkeys,preprintnumbers,amsmath,amssymb]{revtex4}
\usepackage{blkarray, bigstrut} 
\usepackage{graphicx}
\usepackage{graphics}
\usepackage{subfigure}
\usepackage{dcolumn}
\usepackage{bm}
\usepackage{txfonts}
\usepackage{float}
\usepackage{tabu}
\usepackage{xcolor}
\usepackage{color}
\usepackage{makecell}

\usepackage{mathtools}

\begin{document}
\title{Mathematics of multi-agent learning systems at the interface of game theory and artificial intelligence}
\author{Long Wang$^{1,2}$}
\email{longwang@pku.edu.cn}
\author{Feng Fu$^{3,4}$}
\email{fufeng@gmail.com}
\author{Xingru Chen$^{5}$}
\email{xingrucz@gmail.com}
\affiliation{ 
$^1$Center for Systems and Control, College of Engineering, Peking University, Beijing 100871, China \\
$^2$Center for Multi-Agent Research, Institute for Artificial Intelligence, Peking University, Beijing 100871, China \\
$^3$Department of Mathematics, Dartmouth College, Hanover, NH 03755, USA \\
$^4$Department of Biomedical Data Science, Geisel School of Medicine at Dartmouth, Lebanon, NH 03756, USA \\
$^5$School of Science, Beijing University of Posts and Telecommunications, Beijing 100876, China}

\date{\today}

\begin{abstract}
Evolutionary Game Theory (EGT) and Artificial Intelligence (AI) are two fields that, at first glance, might seem distinct, but they have notable connections and intersections. The former focuses on the evolution of behaviors (or strategies) in a population, where individuals interact with others and update their strategies based on imitation (or social learning). The more successful a strategy is, the more prevalent it becomes over time. The latter, meanwhile, is centered on machine learning algorithms and (deep) neural networks. It is often from a single-agent perspective but increasingly involves multi-agent environments, in which intelligent agents adjust their strategies based on feedback and experience, somewhat akin to the evolutionary process yet distinct in their self-learning capacities. In light of the key components necessary to address real-world problems, including (i) learning and adaptation, (ii) cooperation and competition, (iii) robustness and stability, and altogether (iv) population dynamics of individual agents whose strategies evolve, the cross-fertilization of ideas between both fields will contribute to the advancement of mathematics of multi-agent learning systems, in particular, to the nascent domain of ``collective cooperative intelligence'' bridging evolutionary dynamics and multi-agent reinforcement learning.
\end{abstract}



\maketitle


The rapidly advancing era of AI such as the advent of Large Language Models (LLM) has been accompanied by the growing importance of understanding how AI systems interact and make decisions in multi-agent settings. With the surge in the complexity and variety of the interactions and decision-making processes, it is important to develop principled analytical models to understand and further guide these interactions. This perspective will look into the interplay, with an emphasis on the mathematics of multi-agent learning by integrating mathematical principles from game theory and AI. The emerging interdisciplinary research seeks to address multifaceted problems across a spectrum of domains, from the nuances of cultural and social evolution to the challenges of machine behavior in this era of potent AI systems~\cite{zhang2020challenges, tuyls2007evolutionary, bloembergen2015evolutionary, li2022confluence}.

Evolutionary game theory, with its foundations rooted in understanding adaptive and evolving strategies in population dynamics settings, has increasingly become a significant mathematical framework for investigating a wide range of ecological, biological, and social challenges. In the field of biological sciences, researchers have made seminal discoveries regarding the dynamics of host-parasite interactions, the evolution of metabolic pathways, the competition among viruses, and the complexity of cooperation and conflict seen across the natural world. In tackling these questions, the principles of evolutionary game theory find potent applications, providing mechanistic insights to otherwise seemingly puzzling phenomena. Similarly, in the field of social sciences, the backdrop of our socioeconomic systems and human interactions often revolves around strategies, payoffs, and the overarching quest for equilibrium and rich dynamical behavior. How do groups cooperate? Why do certain strategies dominate others? These questions and beyond can be addressed through the lens of game theory~\cite{wang2007evolutionary}.

However, as AI systems permeate our daily lives, a new research frontier emerges. These systems, designed to learn and adapt, increasingly function in heterogeneous environments populated by other AI agents or a mix of humans and machines. How do they make decisions? How do they compete or cooperate? How can we ensure that these AI agents, often driven by ``model-free" complex algorithms with billions of parameters, align with human values and social norms? These are the questions that await our considerations at the confluence of game theory and AI. It is imperative that we endeavor to explore this confluence, focusing particularly on the mathematics of multi-agent learning~\cite{wang2023interdisciplinary}. Our motivation stems from the belief that game-theoretic principles, when integrated with AI, can present a transparent, rigorous, and analytical approach to complex decision-making scenarios in hybrid AI-human systems. This is especially true in contexts such as \textbf{evolutionary computation}, where solutions evolve over time, and \textbf{machine behavior}, where AI systems interact among themselves and with the outside changing environment.

A central aim of integrating EGT and AI will be to establish foundational theories and methodologies that unify these two domains, which will allow for better understanding, prediction, and guidance of AI behavior in multi-agent settings. Additionally, with the growing interactions between humans and intelligent agents in shared environments, there is a pressing need to understand hybrid human-AI systems in various domains, including, but not limited to, medical diagnosis and treatment planning in healthcare, autonomous vehicles with human monitoring in transportation, automated trading with human oversight in finance, and chatbots working alongside human agents in customer service. As such, collaborative efforts across disciplines will extend into developing theoretical and empirical approaches for these hybrid systems. By doing so, we aim to ensure that AI decisions and actions are interpretable, predictable, and synergistic with human intention and moral values. 

To realize these objectives, we need to focus on the development of comprehensive mathematics of multi-agent learning that captures the complexity of multi-agent interactions at the interface of game theory and AI. The corresponding models will seamlessly integrate evolutionary strategies derived from game theory with the cutting-edge learning algorithms representative of AI. A salient feature of the integration will be its emphasis on the nuances of strategies that agents employ in anticipation of recurring interactions, as elucidated through the lens of repeated games and machine-learning-based strategies. For example, a potential testbed for this synergistic integration is the iterated Prisoner's Dilemma, where machine learning methods like particle swarming optimization (PSO) and finite state machine (FSM) are already used to generate sophisticated strategies~\cite{harper2017reinforcement}. While these strategies, particularly those based on neural networks, may exhibit similarities to simple reactive strategies from evolutionary game theory, such as Tit-for-Tat and Win-Stay, Lose-Shift, they can also possess much greater subtlety and nuance in their understanding of the interaction environment, owing to their adaptive reinforcement learning capacities.

Having these foundational models instituted, the subsequent inquiry will involve scrutinizing their implications in a variety of scenarios. To be more specific, both competitive and cooperative interactions (or more generally mixed mode) will be considered to delineate equilibrium states, discern dominant strategies, and pinpoint potential pitfalls. An essential facet of the investigation will center on \textbf{stochastic games} and \textbf{reinforcement learning}, where the unpredictability inherent in agent interactions is juxtaposed with how reinforcement learning dynamics enable agents to reason and infer optimal strategies in these dynamical settings. It is worth noting that the resulting learning equilibria, or more generally, rich dynamical behavior from multi-agent reinforcement learning, can be different from the well-established equilibrium concepts such as Nash equilibrium and evolutionarily stable strategy (ESS). In the interim, understanding AI behavior and the moral characters they exhibit in the context of competition and cooperation should be carried out through the principles of evolutionary game theory. Further in-depth investigation will assess how AI agents evolve their strategies in response to the shifting landscape of their environment and game dynamics. Moreover, it will critically analyze the common characteristics (or disparity) between the evolutionary strategies of AI agents and the moral values and ethics intrinsic to human society.

Furthermore, in the domain of \textbf{cooperative AI systems}, finding common ground via cooperative social norms is crucial~\cite{dafoe2021cooperative}. The underlying rationale is crystal clear: for the effective and ethical deployment of AI techniques in human-driven ecosystems, fostering and upholding cooperative systems is paramount. A primary objective is to study the mechanisms that steer AI agents towards cooperation, commonality, and the establishment of universally beneficial social norms. 

To this end, the first thing we need to consider is to examine not only the technical aspects of AI systems but also the ethical concerns underpinning their cooperative behavior. In the pursuit of an immersive integration into societal frameworks, it is crucial to understand how to infuse AI agents with a sense of ethical responsibility and alignment with human values. Doing so requires not just programming algorithms for efficiency but also instilling a deeper comprehension of social norms that resonate with human perspectives. 

A second thing we should keep in mind is that the study of cooperative AI systems will one day extend to the development of adaptive models that can respond dynamically to the ever-changing subtleties of human interactions. The introduction of new social media platforms and advanced communication tools can alter the dynamics of both public and private interactions, and shifts in cultural attitudes such as those related to gender equality, environmental awareness, and civil rights, for instance, can result in cultural shifts and social movements. As social dynamics unfold and ethical norms evolve, AI systems must be equipped with the flexibility to adapt while upholding cooperative principles, thereby ensuring that human-made agents remains a supportive force in human environments.

\begin{figure*}[htbp]
\centering
\includegraphics[width=0.9\textwidth]{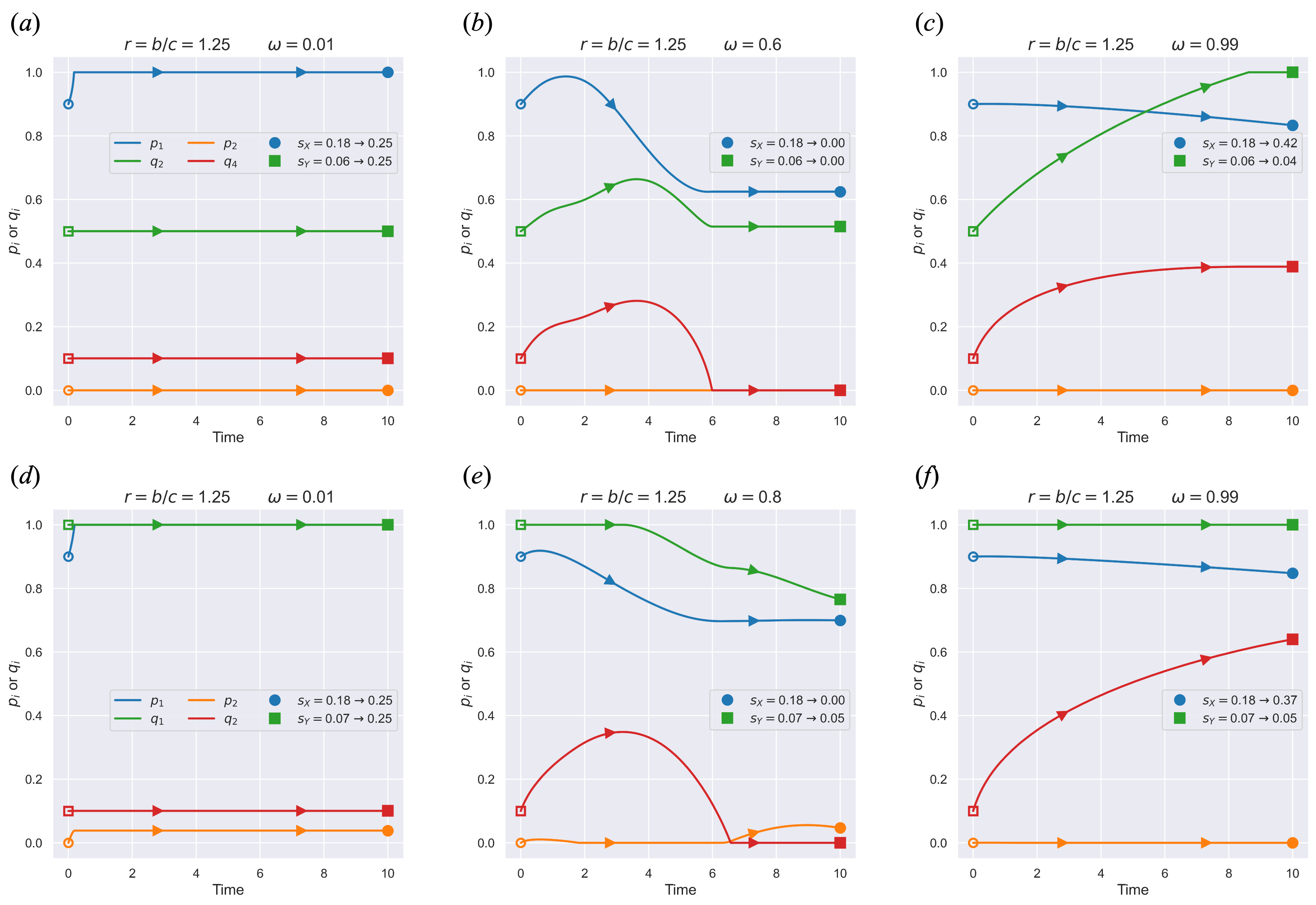} \\
\caption{Leveraging unbending strategies for steering and stabilizing fairness and cooperation. Shown is co-adaptive learning dynamics between two players in a repeated donation game. The initial strategy used by player X is an extortionate zero-determinant strategy $[p_1, p_2, p_1, p_2]$ with extortion factor $\chi = 2.8$ and that used by player Y is (a)-(c) an unbending strategy $[1, q_2, 0, q_4]$ from Class A or (d)-(f) an unbending strategy $[q_1, q_2, q_1, q_2]$ from Class D (see Ref.~\cite{chen2023outlearning} for detailed classifications). The benefit-to-cost ratio of the donation game is referred to as $r$ and the relative time scale governing the time evolution of the behavioral change of player Y as compared to player X is denoted by $\omega$. In light of both players aiming to maximize their respective payoffs $s_X$ and $s_Y$, the learning curves of (a)-(c) $p_1$, $p_2$, $q_2$, and $q_4$ or those of (d)-(f) $p_1$, $p_2$, $q_1$, and $q_2$ are shown with respect to time. The initial and final payoffs of the two players are given. The circles and the squares stand for player X and player Y, the empty and the solid points represent the initial and the final states, and the arrows indicate the directions of learning. For comparison, three different cases are considered: (a) (d) $\omega \to 0$, (c) (f) $\omega \to 1$, and (b) (e) $\omega$ in between.}
\label{fig1}
\end{figure*}

In the endeavor towards understanding \textbf{multi-agent learning systems}, we need to emphasize consensus and convergence. A particular phenomenon here worthy of highlighting is the \textbf{Red Queen dynamics}. Namely, the evolutionary arms race between learning agents, characterized by their nontrivial co-adaption and co-evolution of strategies to gain a competitive edge, calls for a deep exploration~\cite{chen2023outlearning}. An example of this sort arising in the context of repeated interactions is presented in Figure~\ref{fig1}. We consider a simplified Prisoner's Dilemma, known as the donation game, and observe adaptive learning players who aim to maximize their payoffs. To account for a tug-of-war situation in the adaptive dynamics of behavior response, we introduce a relative time scale $\omega$ that governs the time evolution of the behavioral change of the second player Y as compared to the first player X. For small values of $\omega$, in particular, $\omega \to 0$, the resulting dynamics reverts to the steering learning scenario where player Y is fixed in their behavior. In contrast, as $\omega \to 1$, the dynamics converges to the scenario features a fixed player X. For intermediate values of $\omega$, an interesting arms race can emerge between the two adaptive players, and neither of them gains a long-term increase in payoff.

The ultimate cornerstone of the related interdisciplinary research agenda in this regard will be the establishment of alignment mechanisms with a noble vision of serving all humanity. Towards this goal, \textbf{collective cooperative intelligence} will help play a pivotal role in fostering collaboration, innovation, and shared efforts toward the betterment of all beyond scientific research~\cite{leonard2022collective}. As shown in biological, social, and organizational contexts, individuals can collaborate and work together to achieve the greater good. A similar synergy can be envisioned among intelligent agents, where their individual contributions complement each other, resulting in enhanced problem-solving capabilities and decision-making outcomes that surpass what each agent could achieve alone. The growing idea of \textbf{collective reinforcement learning dynamics}, designed to combine the analytical tractability of evolutionary dynamics and the intrinsic complexity of reinforcement learning, thus serves as a concrete catalyst for making promising and impactful progresses. It will enable the appealing capacity to explore complex dynamics in a multi-agent system yet with reduced computational cost and enhanced efficiency compared to traditional reinforcement learning algorithms.

In conclusion, we foresee promising, productive, and fruitful adventures at the emerging frontier where evolutionary game theory meets the challenges as well as the promises of artificial intelligence. By integrating these two versatile yet powerful domains, we expect insightful findings and results that help unravel the nuances and subtleties in hybrid AI-human systems. These integrative efforts will provide AI scholars, practitioners, and enthusiasts with a comprehensive guide that demystifies the convergences and, in a positive light, establishes a robust academic foundation. In particular, it will offer a unified perspective on the evolving landscapes of game theory and artificial intelligence. In the not-so-distant future, we, as a diverse community of researchers who aspire to push the boundaries of our understanding, will together lay the groundwork for the next generation of multi-agent AI systems. The collective outcomes of this kind will not only hold academic significance but will also have practical implications, shaping the future coexistence of humanity with AI in an ethical and beneficial manner.


\section*{Acknowledgements.}
This work is supported by the National Natural Science Foundation of China (grant no.~62036002) and Beijing Natural Science Foundation (grant no.~1244045).

\nocite{*}
\bibliography{ref}

\end{document}